\documentclass[letterpaper,twocolumn,prl,aps,superscriptaddress,floatfix,10pt]{revtex4-2}
\usepackage{amsmath}
\usepackage{amssymb}
\usepackage{amsfonts}
\usepackage{graphicx}
\usepackage{braket}
\usepackage[colorlinks, citecolor=red]{hyperref}
\usepackage{titletoc}

\begin{document}

\title{Global Parametric Gates for Multi-qubit Entanglement}
\author{J.Z.Yang}
\affiliation{Center for Quantum Information, Institute for Interdisciplinary Information Sciences, Tsinghua University, Beijing 100084, China}

\author{L.Guo}
\affiliation{Center for Quantum Information, Institute for Interdisciplinary Information Sciences, Tsinghua University, Beijing 100084, China}

\author{H.N.Xiong}
\affiliation{Beijing Key Laboratory of Fault-Tolerant Quantum Computing, Beijing Academy of Quantum Information Sciences, Beijing 100193, China}

\author{J.H.Wang}
\affiliation{S Lab at Shenzhen’s Quantum Science Center in the Guangdong Hongkong-Macao Greater Bay Area, 518045, China}

\author{Y.Li}
\affiliation{Center for Quantum Information, Institute for Interdisciplinary Information Sciences, Tsinghua University, Beijing 100084, China}

\author{Y.F.Yang}
\affiliation{Center for Quantum Information, Institute for Interdisciplinary Information Sciences, Tsinghua University, Beijing 100084, China}

\author{C.J.An}
\affiliation{Center for Quantum Information, Institute for Interdisciplinary Information Sciences, Tsinghua University, Beijing 100084, China}

\author{H.Y.Zhang}
\affiliation{Center for Quantum Information, Institute for Interdisciplinary Information Sciences, Tsinghua University, Beijing 100084, China}
\affiliation{Hefei National Laboratory, Hefei 230088, China}

\author{L.Y.Sun}
\affiliation{Center for Quantum Information, Institute for Interdisciplinary Information Sciences, Tsinghua University, Beijing 100084, China}
\affiliation{Hefei National Laboratory, Hefei 230088, China}

\author{Y.P. Song}\email{ypsong@mail.tsinghua.edu.cn}
\affiliation{Center for Quantum Information, Institute for Interdisciplinary Information Sciences, Tsinghua University, Beijing 100084, China}
\affiliation{Hefei National Laboratory, Hefei 230088, China}

\author{L.M.Duan}\email{lumingduan@tsinghua.edu.cn}
\affiliation{Center for Quantum Information, Institute for Interdisciplinary Information Sciences, Tsinghua University, Beijing 100084, China}
\affiliation{Hefei National Laboratory, Hefei 230088, China}

\begin{abstract}
We propose and experimentally demonstrate a global parametric gate that generates multi-qubit entangled states in a single step. 
By applying a parametric drive to a common qubit at precise detunings relative to computational qubits, we directly produce two-, 
three-, and four-qubit entanglement with state fidelities of 99.4\%±0.2\%, 93.4\%±0.3\%, and 91.4\%±0.3\%, respectively. This scheme 
enables efficient, reconfigurable control using only microwave drives and is compatible with fixed-frequency qubits. Error analyses indicate 
that infidelity stems primarily from decoherence and coherent control errors, with negligible contributions from 
static ZZ coupling and flux noise. Furthermore, simulations with state-of-the-art parameters predict this global gate can generate 
high-fidelity (99.70\%) entanglement in systems of up to six qubits.
\end{abstract}

\maketitle

All quantum information technologies rely on a fundamental prerequisite: high-fidelity quantum operations. This includes 
high-fidelity gates\cite{sung2021realization,li2024realization,ding2023high,kim2022high}, on-demand generation of entangled states
\cite{song201710,reagor2018demonstration,gong2019genuine,friis2018observation}, and precise coherent control
\cite{barends2014superconducting,arute2019quantum,boixo2018characterizing,landsman2019verified,wang2022information}. Simulating quantum many-body 
systems is a key application for emerging quantum processors\cite{fauseweh2024quantum,o2016scalable,roushan2017spectroscopic,maskara2025programmable}. 
As superconducting processors scale toward increasingly 
complex qubit networks, two critical challenges emerge: developing fault-tolerant architectures resilient to component 
failures\cite{versluis2017scalable,chamberland2020topological}, and significantly reducing the resource overhead for qubit control
\cite{li2020towards,gambetta2017building,cross2019validating}.

State-of-the-art superconducting quantum architectures currently employ either one-dimensional\cite{barends2014superconducting,roushan2017spectroscopic,kelly2015state}
or heavy-square lattice configurations\cite{arute2019quantum,versluis2017scalable,takita2017experimental,gong2021quantum}, 
where each transmon qubit incorporates independent XY and Z control lines for gate operations and 
frequency tuning. While these platforms have successfully demonstrated quantum advantage and enabled surface code 
error correction implementations, they remain vulnerable to network fragmentation from individual qubit failures. 
In addition, long-range interactions in many-body systems can give rise to rich physical phenomena and complex 
phase diagrams\cite{defenu2023long,lyu2023variational}. 
However, conventional superconducting quantum architectures typically restrict interaction distances. 
A ring-network superconducting architecture features shared bus resonators - each ring contains a common bus resonator 
enabling all-to-all connectivity among qubits. By leveraging engineered long-range interactions and all-to-all connectivity, 
the ring structure inherently prevents network fragmentation from individual qubit failures, and thus enhances 
quantum processor robustness. 

Quantum entanglement is a hallmark of quantum systems, distinguishing them from classical ones\cite{gong2019genuine,friis2018observation}. 
However, preparing large-scale multi-qubit entangled states remains challenging, as it requires complete control over the qubits and 
their noise environment. Multi-qubit entanglement has been demonstrated in platforms including superconducting 
circuits and ion traps\cite{song201710,reagor2018demonstration,gong2019genuine,friis2018observation,wang201816}. 
Conventional methods typically employ sequential two-qubit entangling gates, an approach 
whose scalability is constrained by qubit coherence. Exploiting global quantum gates that generate multi-qubit 
entanglement in a single-step would help reduce quantum circuit depth and improve simulation efficiency. Theoretical 
predictions suggest that global multi-qubit entangling gates with all-to-all connectivity could provide the 
key to achieving polynomial or even exponential speedups\cite{lu2019global,maslov2018use,martinez2016compiling}. 
While parallel operations on five qubit pairs have 
been used to generate Greenberger-Horne-Zeilinger (GHZ) states\cite{song201710}, this method requires all qubits to be equally 
coupled to and detuned from a bus resonator, imposing fabrication challenges and necessitating flux tunability.

Furthermore, the gate scheme is expected to be compatible with fixed-frequency qubits with microwave-only control, 
reducing the resource overhead for qubit control. Such schemes have been investigated, including cross-resonance (CR) 
gate\cite{sheldon2016procedure,cai2021impact}, resonator-induced phase gate (RIP)\cite{paik2016experimental}\textcolor{red}{,}
and proposed geometric phase gates\cite{cai2021all,zhu2003unconventional}. While the CR gate does not support 
a global gate scheme, the refocused RIP gate could realize a global entangling gate\cite{paik2016experimental}, but only if each qubit's 
frequency is individually tunable or by employing an echo trick. In this letter, we demonstrate a global 
entangling gate in a ring-network superconducting processor with a high-connectivity architecture, using a bus resonator 
for programmable inter-qubit coupling. By parametrically driving on a common qubit with precise frequency 
detunings relative to the computational qubits, we generate a four-qubit entangled state in a single step with 
a fidelity of 91.4\% (±0.3\%). This unique global gate scheme excels at producing multi-qubit entanglement efficiently, 
offering reconfigurable control solely through parametric microwave drives. The method enables the use of fixed-frequency 
computational qubits, providing frequency selectivity while alleviating frequency crowding and minimizing sensitivity 
to environmental noise.

\begin{figure}[t]
\includegraphics[width=0.5\textwidth]{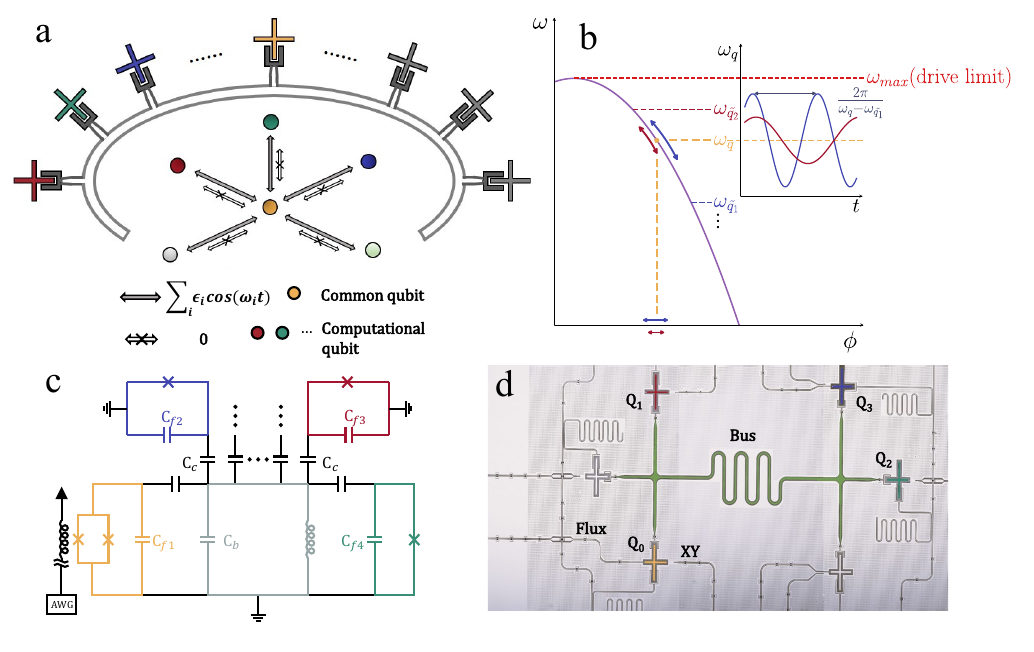}
\caption{Device architecture and gate principle. (a-b) Working principle of the global parametric gate. 
          Multiple computational qubits (red, blue, green…) couple to a shared bus resonator, enabling all-to-all connectivity. 
          Entanglement across multiple qubits is generated in a single step by applying a microwave parametric drive to a flux-tunable 
          common qubit (yellow). The drive consists of a superposition of frequency components detuned from the fixed-frequency 
          computational qubits. (c) Schematic circuit of the superconducting architecture featuring the shared bus resonator. 
          The parametric drive is applied to the common qubit via its dedicated flux line. (d) False-colored optical image of the fabricated device.}
\label{fig:Fig1}
\end{figure}

\emph{Working principle of global parametric gate.}
\autoref{fig:Fig1}\textcolor{red}{(a),(b)} illustrates the working principle of our global parametric gate. Here, the frequency-tunable common qubit 
is biased at an operational point, where the frequency deviation from the sweet spot is larger than its detuning 
from each computational qubit. We generate an effective coupling $g_{j,eff}$, between the common qubit and the j-th 
computational qubit mediated by the bus resonator, by applying a microwave drive to the common qubit's flux line. 
This drive is a superposition of the specific frequency components that detune the common qubit from each computational 
qubit, given by $\sum_j \Omega_j \sin{(\Delta_j t+\phi_j)}$ , where $\Delta_j=\omega_j-\omega_0$ is the frequency detuning between the common 
qubit ($\omega_0$) and the j-th computational qubit ($\omega_j$). The coupling $g_{j,eff}$ can be tuned independently by modulating 
the amplitude $\Omega_j$ and phase $\phi_j$ of each drive frequency component. We initialize all qubits to the ground state $\ket{0}$. 
A $\pi$ pulse is then applied to one qubit to prepare the initial state $\ket{1}\ket{0}^{\otimes j-1}$. Subsequently, the parametric 
drive is activated for a duration t, satisfying the condition $g_{j,eff}t = (n+\frac{1}{4})\pi$ for an integer n. This operation generates 
an j-qubit W state: $\ket{W_j}=\frac{1}{\sqrt{j}}(\ket{100\cdots0}+\ket{010\cdots0}+\cdots+\ket{00\cdots01})$, under the condition 
that the coupling strength $g_{j,eff}$ is identical for all computational qubits to the common qubit. 
This gate scheme differs from conventional parametric approaches, which either use a microwave-driven 
tunable coupler for two-qubit gates\cite{li2018perfect,mckay2016universal,han2020error} or generate multi-qubit states 
by sequentially applying two-qubit gates to qubit pairs\cite{reagor2018demonstration}.

\emph{Experimental setup.} Our quantum processor consists of six transmon qubits ($Q_0$-$Q_5$) coupled to a shared bus resonator. For 
implementation of the global entangling gate, we use qubits $Q_0$ as the common qubit and $Q_1$-$Q_3$ as the computational 
qubits, as illustrated in \autoref{fig:Fig1}(d). All four qubits are negatively 
detuned from the bus resonator, with qubit frequencies $\omega_j/2\pi$ = 5.0408, 5.0992, 5.2056, and 5.2347\,GHz for j=0,1,2,3, 
respectively, and a resonator frequency $\omega_b/2\pi$=5.5208\,GHz. The measured coupling strengths between each qubit $Q_j$ and the bus resonator— $h_j/2\pi$=17.3, 22.8, 18.2,  
and 15.2 MHz — satisfy the dispersive condition $h_j\ll|\Delta_{j,b} |$, where $\Delta_{j,b}=\omega_j-\omega_b$.

The system Hamiltonian under the rotating-wave approximation (RWA) is given by (see Supplementary Material for details):
\begin{equation}
  \begin{aligned}
  H=&\sum_{j=0}^n \frac{\alpha_j}{2} a_j^\dagger a_j^\dagger a_j a_j + \sum_{j=1}^n g_j (a_0 a_j^\dagger \mathrm{e}^{i\Delta_j t}+h.c.)\\ 
    &+\sum_{j=1}^n \Omega_j \sin{(\nu_j t+\phi_j)} a_0^\dagger a_0
  \end{aligned}
\end{equation}
Where $g_j=\frac{h_0h_j}{2}(\frac{1}{\Delta_{0,b}}+\frac{1}{\Delta_{j,b}})$, $\Omega_j$  and $\nu_j$  are the parametric drive amplitude 
and frequency for the qubit pair $Q_0$-$Q_j$. $\alpha_j$ denotes the anharmonicity of $Q_j$. 

In the interaction picture, under the unitary transformation $U=\mathrm{Exp}\{i\sum_{j=1}^n \epsilon_j \cos{(\nu_j t+\phi_j)a_0^\dagger a_0}\}$, 
the Hamiltionian becomes:
\begin{equation}
 H_I = \sum_{j=1}^n g_j \left(a_0 a_j^\dagger  \mathrm{e}^{i\Delta_j t}\mathrm{e}^{\sum_{j=1}^n \epsilon_j \cos{(\nu_j t+\phi_j)}} + h.c.\right)
\end{equation}
where $\epsilon_j = \Omega_j/\nu_j$. When $\nu_j=\Delta_j$ and the effective drives, $\epsilon_1,\epsilon_2,...,\epsilon_j$, are identical, we obtain 
the final effective Hamiltionian, 
\begin{equation}
  H = \sum_{j=1}^n g_j J_1 (\epsilon_j ) \prod_{k\neq j} J_0 (\epsilon_k) (i a_0 a_j^\dagger \mathrm{e}^{i\phi_j}+h.c.)
\end{equation}

This effective Hamiltonian describes a controllable exchange interaction with an effective coupling strength 
$g_{j,eff}=g_jJ_1(\epsilon_j)\prod_{k\neq j} J_0 (\epsilon_k) $, where $J_n$ are Bessel functions of the first kind\cite{li2018perfect,wu2018efficient}.

We begin by characterizing the two-qubit parametric gate. The operating frequency of the common qubit, $Q_0$, 
is set to 5.0408\,GHz by applying a specific bias voltage to its flux line (see Supplementary Material). 
To account for the nonlinear frequency response, we modulate the parametric drive, causing the qubit frequency 
oscillates symmetrically about its operational setpoint as $\Omega_j\sin{(\nu_j t+\phi_j)}$.  

\begin{figure}[t]
	\includegraphics[width=0.5\textwidth]{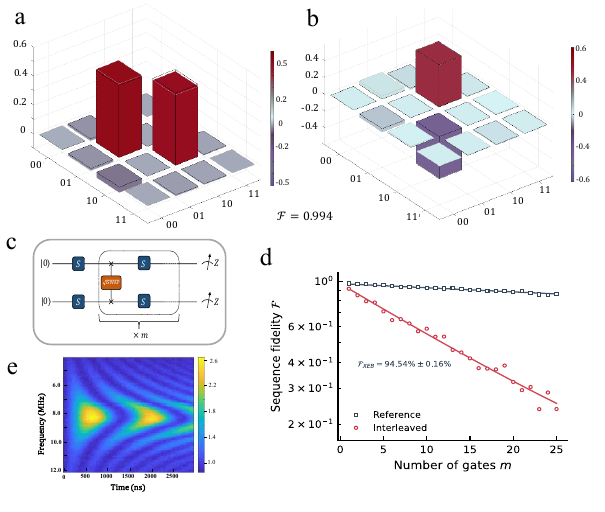}
	\caption{Real part (a) and imaginary part (b) of a reconstructed 2-qubit density matrix for $Q_0$-$Q_1$ 
  (F = 0.994(2), $t_{gate}=320\,\mathrm{ns}$). Experimental results (colored bars) are compared to the ideal 
  state (transparent bars). (c) Circuit schematic of the XEB sequence for the $\sqrt{\mathrm{iSWAP}}$gate. The single-qubit 
  gates S are randomly chosen from a set of 64 distinct single-qubit gates. (d) Fidelity decay for the $\sqrt{\mathrm{iSWAP}}$
  gate ($t_{gate}=316\,\mathrm{ns}$) and the reference gates as a function of the circuit depth m. (e) The exchange oscillation between 
  $\ket{10}$ and $\ket{01}$ for $Q_0$-$Q_1$ as a function of drive length and drive frequency $\nu_{1,\Omega_1}$ relative to the qubit frequency detuning $\Delta_1$ at $\Omega_1=0$. The parametric modulation induces a ~8.3 MHz DC shift in $Q_0$'s frequency, offsetting the resonant exchange from $\nu_{1,\Omega_1=0}$.}
	\label{fig:Fig2}
\end{figure}

\begin{figure}[t]
	\includegraphics[width=0.5\textwidth]{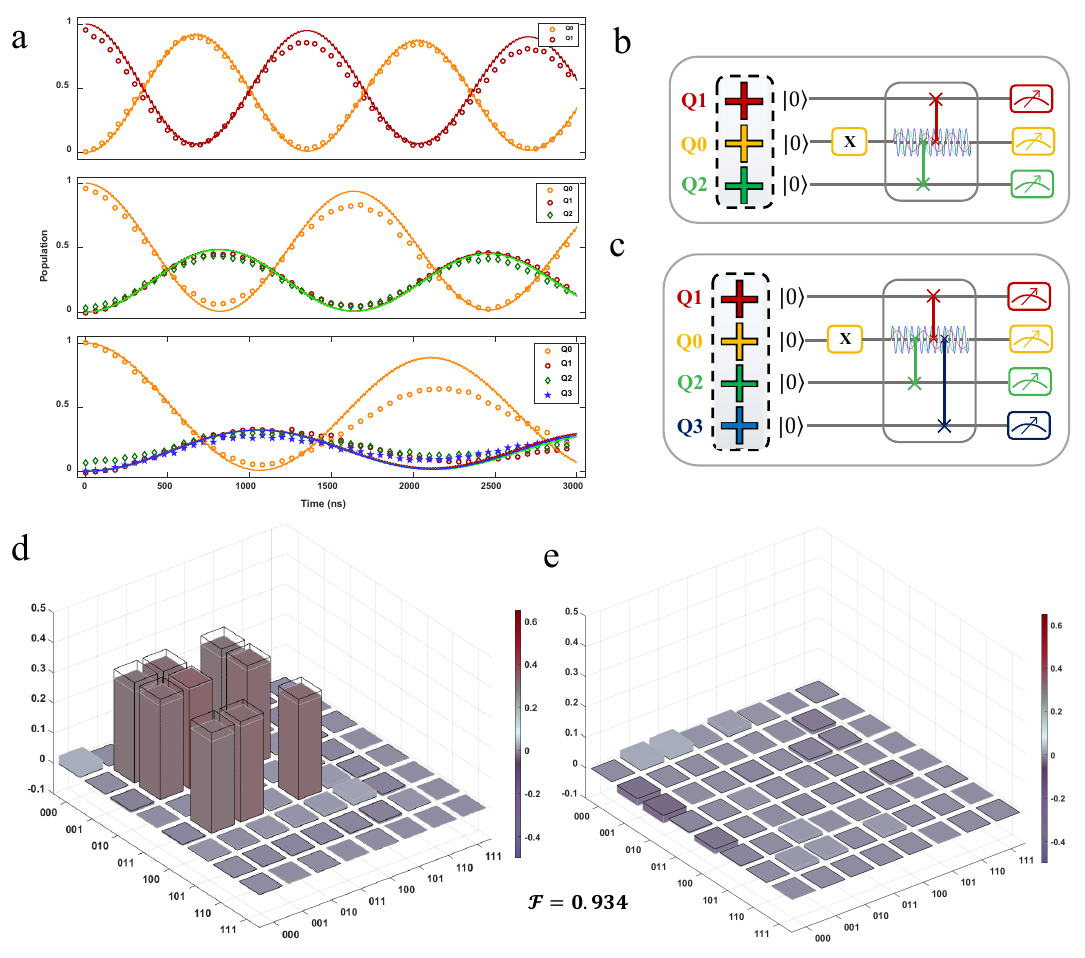}
	\caption{(a) Qubit occupation probabilities during multi-qubit exchange oscillations. The upper, middle, and lower panels correspond to the two-qubit 
  ($Q_0$-$Q_1$, initial state $\ket{01}$), three-qubit ($Q_0$-$Q_1$-$Q_2$, initial state $\ket{100}$), and four-qubit ($Q_0$-$Q_1$-$Q_2$-$Q_3$, initial state $\ket{1000}$) systems, 
  respectively. Experimental data are shown as colored markers, with simulation results as solid lines ($Q_0$: yellow, cycle $Q_1$: red, cycle $Q_2$: green, diamond $Q_3$: blue, pentagram). 
  For the two-, three-, and four-qubit cases, the parametric drive frequencies are shifted by 8.3\,MHz, 7.6\,MHz, and 10.8\,MHz, respectively, 
  relative to the detuning between each computational qubit and the common qubit. 
  Simulation parameters: $\Omega_1/2\pi$=41.4\,MHz, $\nu_1/2\pi$=58.4\,MHz ($Q_0$-$Q_1$); $\Omega_1/2\pi$=29.9\,MHz, $\Omega_1$:$\Omega_2$=1:3, $\nu_1/2\pi$=58.4\,MHz, $\nu_2/2\pi$= 164.8\,MHz ($Q_0$-$Q_1$-$Q_2$); 
  $\Omega_1/2\pi$=18.3\,MHz, $\Omega_1$:$\Omega_2$:$\Omega_3$=1:3:4, $\nu_1/2\pi$ = 58.4\,MHz, $\nu_2/2\pi$ = 164.8\,MHz, $\nu_3/2\pi$ = 193.9\,MHz  ($Q_0$-$Q_1$-$Q_2$-$Q_3$). 
  (b, c) Pulse sequences used for quantum state tomography (QST) in three- and four-qubit entanglement experiments, respectively. 
  (d, e) Real and imaginary parts of the reconstructed three-qubit density matrix for qubits $Q_0$-$Q_1$-$Q_2$ (F = 0.934(3), $t_{gate}$=460\,ns). 
  Experimental results (colored bars) are compared to the ideal state (transparent bars).}
	\label{fig:Fig3}
\end{figure}

By applying the parametric modulation pulse on the common qubit, we measure the simultaneous quantum-state population 
of both qubits (e.g., $Q_0$ and $Q_1$) as a function of gate length and drive frequency. The resulting population exchange, 
shown in \autoref{fig:Fig2}\textcolor{red}{(e)}, demonstrates coherent excitation oscillations between the qubits. The parametric flux modulation 
induces a DC shift in the common qubit's frequency\cite{mckay2016universal,han2020error}, which shifts the resonant exchange frequency from its un-driven value, 
$\nu_{j,\Omega_j=0}$, by approximately 8.3 MHz. In the upper panel of \autoref{fig:Fig3}(a), we plot the exchange oscillation between the two-qubit states 
$\ket{01}$ and $\ket{10}$ at the drive frequency that maximizes the inter-qubit quantum exchange. The oscillation period is determined 
by the effective coupling strength, $g_{j,eff} (\epsilon_j)$. In our experiments, we selected a drive amplitude that maximizes 
this strength, thereby minimizing the oscillation period, as detailed in the Supplementary Material. At specific 
locations in \autoref{fig:Fig3}(a), where $g_{j,eff} t=(n+\frac{1}{4})$, the excitation is equally shared between both qubits, and a 
maximally entangled state $\frac{\sqrt{2}}{2}(\ket{01}+i\ket{10})$ can be generated\cite{mckay2016universal,han2020error}. A pulse length satisfying $g_{j,eff} t=(n+\frac{1}{2})$ 
implements a primitive iSWAP gate, which can be used to construct a universal gate set for quantum computing. By 
applying a parametric drive with a uniform coupling strength $g_{j,eff}$ between the common qubit and all computational 
qubits, we observe exchange oscillations. These oscillations, starting from the initial states $\ket{100}$ and $\ket{1000}$, 
are shown in the middle and lower panels of \autoref{fig:Fig3}(a). The entangled two-, three-, and four-qubit states are generated 
at gate times of 320\,ns, 460\,ns and 720\,ns, respectively, which we verify using quantum state tomography.

\emph{Quantum State Tomography and Cross-entropy Benchmarking.} We characterize the entangled state using quantum state tomography (QST), following the pulse sequence in 
\autoref{fig:Fig3} (b) and (c). Using the $Q_0$-$Q_1$ qubit pair as an example, we initialize both qubits to the ground state $\ket{0}$ and apply 
an $X_\pi$ pulse to $Q_1$. A parametric drive is then activated for 320 ns. Nine different single-qubit operations, 
$\{I,X_{\frac{\pi}{2}},Y_{\frac{\pi}{2}}\}\otimes\{I,X_{\frac{\pi}{2}},Y_{\frac{\pi}{2}}\}$, are applied to the two-qubit 
state right before a joint single-shot readout to perform state tomography. The resulting data, processed with maximum-likelihood 
estimation, yields the 16 Pauli traces used to reconstruct the two-qubit density matrix, $\rho$. The parametric drive induces a 
phase shift on both qubits. To compensate, we apply a virtual Z gate to each qubit to remove the measured phase offset. 
The achieved state fidelities, defined as $F(\rho,\sigma)=\sqrt{\sqrt{\rho}\sigma\sqrt{\rho}}$ , are 99.4\%±0.2\% for $Q_0$-$Q_1$ 
and 99.0\%±0.2\% for $Q_0$-$Q_2$, where $\sigma$ is the density matrix of the ideal final state. \autoref{fig:Fig2} (a) and (b) show a representative two-qubit state density matrix for $Q_0$-$Q_1$. Unless otherwise stated, all state and gate tomography results include post-selection on successful 
ground-state initialization.

To characterize the two-qubit gate while eliminating state preparation and measurement (SPAM) errors, we perform cross-entropy 
benchmarking (XEB)\cite{arute2019quantum,kim2022high}. The protocol, illustrated in \autoref{fig:Fig2}(c), involves applying sequences of random single-qubit rotations chosen 
from a set of 64 distinct single-qubit gates. Each gate is a combination of a $\pi/2$ rotation around an axis in the xy-plane at $n\pi/8$ 
and a Z-rotation of $n\pi/8$ (n = 0,1,...,7). By measuring the output bitstrings {$x_i$} for each gate sequence, we calculate the 
standard XEB fidelity ($F_{XEB}$). 

Precise calibration of the qubit and drive phases, as well as the common qubit's flux control line response, is crucial\cite{johnson2011controlling}. To this end, 
we employ a Cryoscope protocol to characterize the in situ flux, $\phi_{Q_0}$, by measuring the step response of its dedicated flux control 
line at the chip level\cite{rol2020time}. The Cryoscope measurement is a Ramsey-like protocol in which fixed-amplitude square flux pulses of varying 
durations are inserted between two $\pi/2$ pulses. The accumulated phase shift ($\phi$) from the flux pulse reveals the qubit's on-chip 
frequency detuning, calculated as $\Delta \phi/\Delta t$. We convert this detuning into a flux amplitude using the pre-calibrated qubit spectrum, 
yielding the response in a temporal resolution set by the time domain resolution of the flux pulse. To suppress flux ripples and 
correct phase errors, we then apply infinite impulse response (IIR) and finite impulse response (FIR) filters\cite{rol2020time}. Following this 
calibration and optimization, we achieve an average $\sqrt{\mathrm{iSWAP}}$ gate fidelity of 94.54\%±0.16\% via cross-entropy benchmarking (XEB). 
Full methodological details are provided in the Supplementary Material.

\begin{figure}[t]
\includegraphics[width=0.5\textwidth]{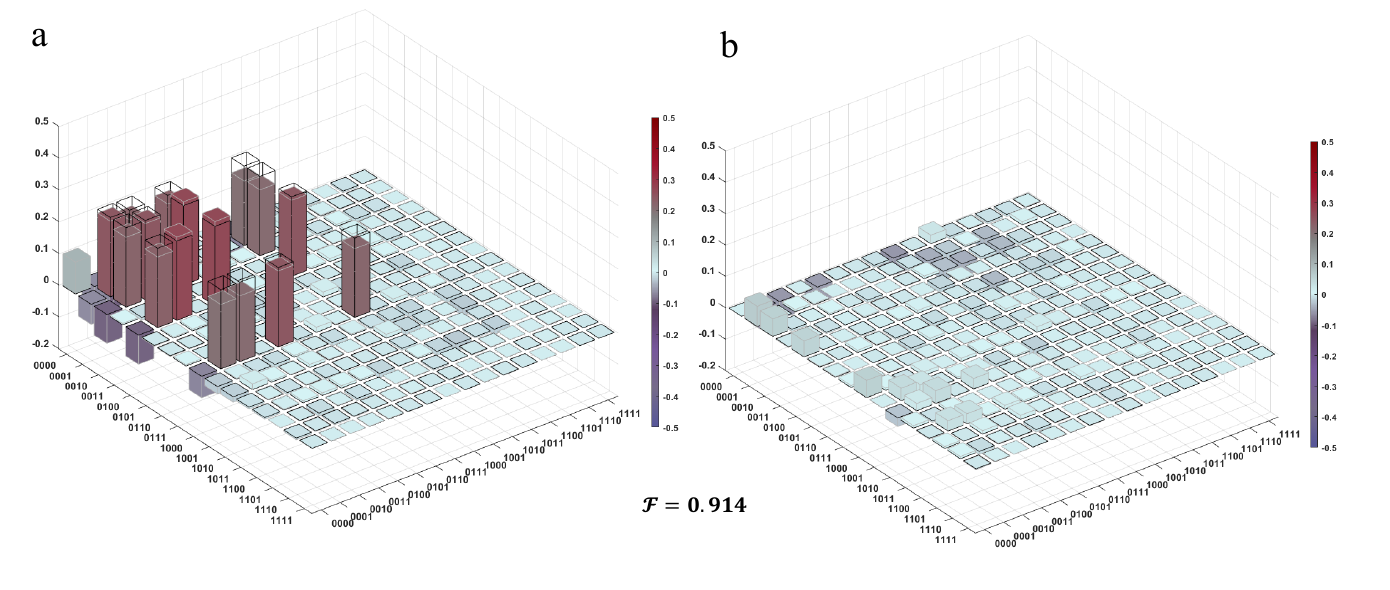}
\caption{Quantum state tomography of four-qubit entanglement. Real (a) and imaginary (b) parts of the reconstructed four-qubit density
 matrix for qubits $Q_0$-$Q_1$-$Q_2$-$Q_3$, measured after a gate time ($t_{gate}$) of 720 ns. The achieved state fidelity is 0.914(3). 
 Experimental results are shown as colored bars, overlaid with the ideal state (transparent bars) for comparison.}
\label{fig:Fig4}
\end{figure}

We now benchmark the multi-qubit global gates using the quantum circuits depicted in \autoref{fig:Fig3} (b) and (c), which are designed to generate entangled 
three- and four-qubit W states, followed by implementing QST on the resulting quantum states. We first calibrate the optimal drive 
amplitude $\Omega_j$ for each parametric component to achieve identical oscillation periods across all qubit pairs. The operational frequency 
detuning and coupling strength vary among the computational qubits and the common qubit. As a result, the three- and four-qubit gate 
times differ from that of a specific two-qubit gate. Similar to the two-qubit operation, we initialize all qubits to the ground state 
$\ket{0}$, apply a $\pi$-pulse to $Q_0$, and then activate the parametric drive for 460\,ns (three-qubit case) and 720\,ns (four-qubit case). 
Measurements in the $X (Y, Z)$ basis are achieved by inserting a Pauli $Y (X, I)$ rotation on each qubit before readout. This yields 
a total of 64 (8 $\times$ 8) and 256 (16 $\times$ 16) Pauli trace datasets for the three- and four-qubit states, respectively, with each dataset 
comprising N = 10000 repetitions. Finally, we reconstruct the density matrix $\rho$ using maximum-likelihood estimation. We achieve state 
fidelities of 93.4\%±0.3\% and 91.4\%±0.3\% for the three- and four-qubit W states, respectively. The representative experimental 
density matrices and ideal matrices are presented in \autoref{fig:Fig3}(d)(e) and \autoref{fig:Fig4}(a)(b).

\emph{Error budget analysis.}
We conduct an error budget analysis to verify the sources of gate error, isolating the contributions from incoherent and coherent noise. 
A quantum error channel can be described by Kraus operators as $\varepsilon(\rho)=\sum_k K_k \rho K_k^\dagger$, 
$\sum_k K_k^\dagger K_k=I$ where $\rho$ and $I$ 
denote the state density and identity operator, respectively. The average fidelity for d-dimensional systems is given by 
$F=\frac{1}{d(d+1)}\sum_k [\mathrm{Tr}(K_k^\dagger K_k)+|\mathrm{Tr}(K_k)|^2]$\cite{li2024realization,ding2023high,pedersen2007fidelity}, 
where $d$ is the dimension of the computational subspace. Based on the Kraus 
operators, the incoherent errors arose from the relaxation ($T_1^q$) and pure dephasing ($T_\phi^q$) of the $q$ th qubit, can be calculated 
as $\alpha \cdot t_g \sum_q(\frac{1}{T_1^q}+\frac{1}{T_\phi^q})$ , with a gate time $t_g$ and a prefactor of 
$\alpha$=2/5,4/9 and 8/17 for the two-, three- and 
four-qubit system, respectively (see Supplementary Material for details)\cite{li2024realization,nielsen2010quantum}. 
Using the qubit coherence time data (see Supplementary 
Material), we calculate coherence-limited infidelities of 1.44\%, 3.35\%, and 7.87\% for the two-, three- and four-qubit gate 
($Q_0$-$Q_1$, $Q_0$-$Q_1$-$Q_2$, $Q_0$-$Q_1$-$Q_2$-$Q_3$), respectively. 

We next consider the flux-noise-induced error from the parametric drive on the common qubit ($Q_0$), which is sensitive to magnetic 
flux fluctuations through its Josephson junction loop. We analyze the $T_2^{Echo}$ data for $Q_0$ using the formula 
$P_{\ket{1}}=A+B\mathrm{e}^{-\Gamma t -(\Gamma_\phi^E t)^2}$, where $\Gamma$ accounts for the effects of energy relaxation 
(T1) and white noise, 
while $\Gamma_\phi^E$ quantifies the dephasing from $1/f$ noise. The flux noise amplitude $\sqrt{A_\Phi}$ is extracted using 
$\Gamma_\phi^E=\sqrt{\ln{2}}|\frac{\partial \omega}{\partial \Phi}|\sqrt{A_\Phi}$, where $S_\phi=A_\Phi / \omega$ and $S_\phi$ is 
the flux-noise spectral density ($\sqrt{A_\Phi}=2.45 {\mu \phi_0})$\cite{sung2021realization,li2024realization,braumuller2020characterizing}. 
Numerical simulations show that the flux-noise-induced incoherent error for the implemented gate (320-720 ns) is on the 
order of $10^{-4}$, which is negligible at the current stage.  

\emph{Numerical simulations.} 
To further investigate potential error sources, we perform numerical simulations using the generalized Hamiltonian of \textcolor{red}{Eq. (S1)}, 
which includes higher-level states. These simulations incorporate qubit decoherence, dephasing, flux noise, and spurious $ZZ$ 
interactions (see Supplementary Material for details). All simulation parameters are set to their experimentally measured values. 
The state fidelity is calculated with $F(\rho,\sigma)=\sqrt{\sqrt{\rho}\sigma\sqrt{\rho}}$  , where $\sigma$ is the 
density matrix of the ideal final state and $\rho$ is the 
simulated density matrix. We estimated the gate fidelity by either simulating the evolution of the gate unitary operator or averaging the state fidelities over 36 initial two-qubit states. 
These states form the set $\{\ket{0},\ket{1},\frac{\sqrt{2}}{2}(\ket{0}\pm\ket{1}),\frac{\sqrt{2}}{2}(\ket{0}\pm i\ket{1})\}^{\otimes 2}$, 
which are generated from the six standard initial single-qubit states.

We first simulate the exchange oscillations for the two-, three-, and four-qubit states; the results are plotted as solid
lines in \autoref{fig:Fig3}(a). The simulated oscillations show good agreement with the experimental data without the use of any tunable 
parameters. Two-qubit quantum state tomography (QST) simulations yield entangled state fidelities of 99.39\% for the $Q_0$-$Q_1$ pair and 99.38\% 
for the $Q_0$-$Q_2$ pair, using the initial states in $\{\ket{0},\ket{1}\}^{\otimes 2}$ These results agree with 
the experimentally measured fidelities. Decoherence and dephasing account for an infidelity of 0.60\% for $Q_0$-$Q_1$ and 0.61\% for $Q_0$-$Q_2$. 
A negligible infidelity of less than $10^{-4}$ is attributed to unwanted static $ZZ$ interaction. Furthermore, 
when comparing the 36 initial states, we find that the fidelities for states in $\{\ket{0},\ket{1}\}^{\otimes 2}$ (99.39\%) are higher 
than those in $\{\ket{+},\ket{i+}\}^{\otimes 2}$(96.57\%). This indicates that the latter states are more susceptible to decoherence/dephasing 
error.

The simulated gate fidelity for pair $Q_0$-$Q_1$ ($Q_0$-$Q_2$) is 98.22\% (97.84\%), with infidelities of 1.53\% (1.72\%) from decoherence/dephasing
and 0.08\% (0.23\%) from the static ZZ interaction. This is higher than the experimental XEB fidelity of 94.54\%±0.16\% ($Q_0$-$Q_1$), 
revealing a discrepancy of ~3.68\% that we attribute to other coherent control imperfections.

Finally, quantum state tomography simulations for the three- and four-qubit entangling states achieve fidelities of 98.99\% and 98.34\%,
respectively. The primary simulated infidelities are 0.99\% and 1.64\%, attributed to decoherence and dephasing, while the static 
$ZZ$ interaction and flux noise are negligible contributors, both on the order of $10^{-4}$. However, the experimentally measured fidelities are lower, at 93.4\%
and 91.4\%. The discrepancies of approximately ~5.59\% and 6.94\% are primarily due to SPAM errors and other coherent control errors, 
which are not included in the simulation. 

\begin{table*}[t]
    \caption{\label{tab:Table 1}
    Simulated error sources for the two-, three-, and four-qubit entanglement. 
            Simulation parameters: 
            1: $\Omega_1/2\pi$=44.6\,MHz, $\nu_1/2\pi$=58.4\,MHz, $t_{gate}$=320\,ns ($Q_0$-$Q_1$); 
            2: $\Omega_2/2\pi$=136.1\,MHz, $\nu_2/2\pi$= 154.9\,MHz, $t_{gate}$=365\,ns ($Q_0$-$Q_1$); 
            3: $\Omega_1/2\pi$=32.6\,MHz, $\Omega_1$:$\Omega_2$=1:3, $\nu_1/2\pi$ = 58.4\,MHz, $\nu_2/2\pi$ = 164.8\,MHz, 
            $t_{gate}$=460\,ns ($Q_0$-$Q_1$-$Q_2$); 
            4: $\Omega_1/2\pi$=18.3\,MHz, $\Omega_1$:$\Omega_2$:$\Omega_3$=1:3:4, 
            $\nu_1/2\pi$ = 58.4\,MHz, $\nu_2/2\pi$ = 164.8\,MHz, $\nu_3/2\pi$ = 193.9\,MHz, $t_{gate}$=720\,ns ($Q_0$-$Q_1$-$Q_2$-$Q_3$); 
            5: $\Omega_1/2\pi$=55\,MHz,$\nu_1/2\pi$ = 59.4\,MHz,$t_{gate}$=316\,ns ($Q_0$-$Q_1$); 
            6: $\Omega_2/2\pi$=135\,MHz, $\nu_2/2\pi$ = 154.0\,MHz, $t_{gate}$=366\,ns ($Q_0$-$Q_2$). 
    Columns show the breakdown of simulated infidelity contributions: 
    $\epsilon_{\text{decoh}}$ (decoherence/dephasing), 
    $\epsilon_{\text{ZZ}}$ (static ZZ interaction), and 
    $\epsilon_{\text{flux}}$ (flux noise).
    }
    % \vspace{-0.5cm}
    \begin{ruledtabular}
        \begin{tabular}{l c c c c c} 
        % I changed 'c' to 'l' (left align) for the first column, usually looks better
            \textrm{Device} & 
            $\mathcal{F}_{\text{sim}}$ & 
            $\epsilon_{\text{decoh}}$ & 
            $\epsilon_{\text{ZZ}}$ & 
            $\epsilon_{\text{flux}}$ & 
            $\mathcal{F}_{\text{exp}}$ \\
            \colrule
            2Q QST ($Q_0$-$Q_1$)    & 99.39\% & 0.60\% & $1.1\times 10^{-5}$ & $1.9\times 10^{-5}$ & $99.4\%\pm0.2\%$\\
            2Q QST ($Q_0$-$Q_2$)    & 99.38\% & 0.61\% & $1.2\times 10^{-5}$ & $2.7\times 10^{-5}$ & $99.0\%\pm0.2\%$\\
            3Q QST ($Q_0$-$Q_1$-$Q_2$) & 98.99\% & 0.99\% & $3.3\times 10^{-4}$ & $3.1\times 10^{-4}$ & $93.4\%\pm0.3\%$\\
            4Q QST (All)      & 98.34\% & 1.64\% & $6.4\times 10^{-4}$ & $7.2\times 10^{-4}$ & $91.4\%\pm0.3\%$\\
            Gate ($Q_0$-$Q_1$)      & 98.22\% & 1.53\% & $8.0\times 10^{-4}$ & $1.2\times 10^{-3}$ & $94.54\%\pm0.16\%$\\
            Gate ($Q_0$-$Q_2$)      & 97.84\% & 1.72\% & $2.3\times 10^{-3}$ & $1.5\times 10^{-3}$ & -- \\
        \end{tabular}
    \end{ruledtabular}
    % \vspace{-0.5cm}
\end{table*}

Based on the above analyses, we conclude that the error in the multiparticle entanglement state generated by the global parametric 
scheme originates primarily from decoherence and coherent control errors. In contrast, the contributions 
from static $ZZ$ and flux noise are negligible, on the order of $10^{-4}$. These error sources and their corresponding budgets are summarized 
in \autoref{tab:Table 1}.

\emph{Discussion and summary.} 
The W state is a valuable resource in quantum information processing due to its robustness and persistent entanglement\cite{park2025entangled}, making it 
suitable for tasks where quantum networks and communication are involved, and where the loss of qubits is a concern. 
Reducing the frequency detuning between the common and computational qubits while enhancing their interaction with the 
bus significantly shortens gate time, thereby suppressing decoherence error. For example, simulation shows that a $\sqrt{\mathrm{iSWAP}}$ 
gate can be realized in 50 ns with a frequency detuning of 100 MHz ($\Delta_1/2\pi$=100.0\,MHz, $\omega_b/2\pi$=5.1922\,GHz) and a qubit-bus 
coupling of 30\,MHz ($h_{0,1}/2\pi$=30\,MHz), with all other parameters matching those in \autoref{fig:Fig3}(a). Using state-of-the-art superconducting qubit parameters 
(see Supplementary Material), we simulate the generation of entangled states via parametric global gates in two- to six-qubit systems, accounting 
for decoherence, static ZZ coupling, and flux noise. The resulting state fidelities are 99.91\%, 99.88\%, 99.85\%, 99.80\%, and 99.70\%, 
respectively, well above the threshold for genuine 
multipartite entanglement of W states\cite{lougovski2009verifying}. This implies that within a hexagonal ring-network lattice, high-fidelity intra-ring 
multipartite entanglement can be generated in a single step. Furthermore, by parametrically driving the common qubit shared 
between adjacent rings, we can also achieve multi-qubit entanglement across the lattice. This scheme thus enables long-range 
interactions and creation of multipartite entangled states across a honeycomb superconducting lattice.

\bigskip
\section*{Acknowledgements}
We thank Haiyan Wang and Yue Jiang for the technical support.This work is supported by Quantum Science and Technology-National Science and Technology Major Project, Grant No. 2021ZD0301700.

\bigskip
\section*{Author Contributions}
L.M.D. and Y.P.S supervised the project. Y.P.S and J.Z.Y performed the measurements and simulations. Y.P.S fabricated the device. Y.P.S wrote the 
manuscript with feedback from all authors, and J.Z.Y. wrote specific sections of the supplementary materials.

\bigskip
\section*{Competing interests}
The authors declare no competing interests.

\bibliographystyle{apsrev4-2} 
\bibliography{all_ref} % Points to my_references.bib

\clearpage % 换页
\phantomsection 
\pdfbookmark[1]{Supplemental Material}{SM_Cover}
\onecolumngrid % 如果正文是双栏，SM 通常建议单栏（可选）

% --- 开始 Supplementary Material ---
\begin{center}
\textbf{\large Supplementary Material}
\end{center}

% 重置计数器，让图表公式编号变为 S1, S2...
\setcounter{equation}{0}
\setcounter{figure}{0}
\setcounter{table}{0}
\setcounter{page}{1}
\makeatletter
\renewcommand{\theequation}{S\arabic{equation}}
\renewcommand{\thefigure}{S\arabic{figure}}
\renewcommand{\thetable}{S\arabic{table}}
\makeatother

\startcontents[sm]
\begingroup
\makeatletter
% Fix for titletoc and revtex4-2 conflict
\def\l@section{\@dottedtocline{1}{1.5em}{2.3em}}
\def\l@subsection{\@dottedtocline{2}{3.8em}{3.2em}}
\makeatother
\printcontents[sm]{}{1}{\section*{Contents}}
\endgroup

% \startcontents[sm]
% \printcontents[sm]{}{1}{\section*{Contents}}

\section{Device fabrication}
    The device fabrication involves three primary steps: (1) A c-plane sapphire substrate is 
    chemically cleaned with piranha solution, annealed, and then coated with a tantalum (Ta) 
    film via sputtering deposition. Photolithography and inductively coupled-plasma (ICP) 
    etching are subsequently used to define all base wiring and resonators. (2) Two photolithography 
    processes, followed by Ta deposition and wet etching, are employed to construct air bridges. 
    These structures primarily serve to connect segments of ground plane, thereby reducing 
    parasitic slotline modes. (3) Josephson junctions are fabricated using electron-beam (e-beam) 
    lithography with two layers of e-beam resists (A4/EL11) and a double-angle aluminum deposition process. 
    Prior to spin-coating the e-beam resists, the chip undergoes sequential treatment in TAMI, 
    piranha solution, and buffered oxide etch (BOE). We optimized the sputtering conditions—including 
    pressure, deposition rate, substrate temperature, and target-to-substrate distance—to produce 
    high-quality alpha-phase tantalum films.
\section{Measurement setup}
\begin{figure}[!ht]
        \centering
        \includegraphics[width=0.8\textwidth]{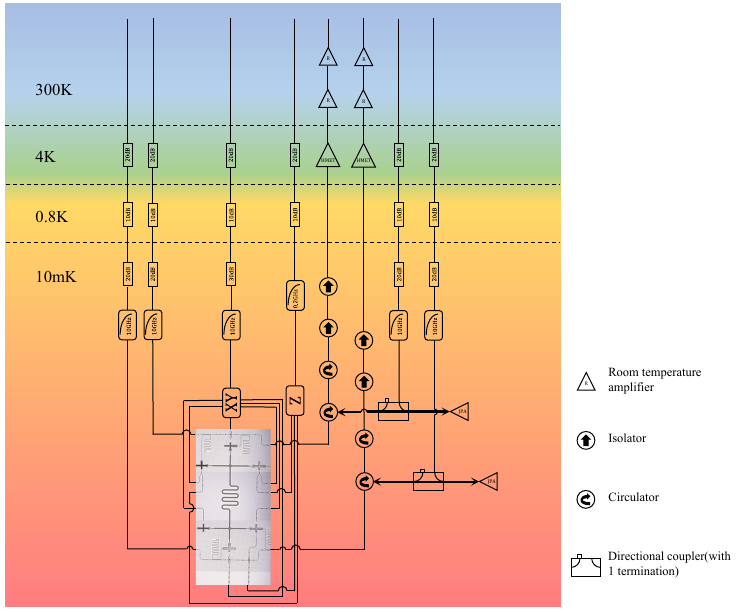}
        \caption{ \textbf{Experimental setup and cryogenic wiring diagram}. The schematic illustrates the signal path 
        for qubit control and measurement. Control signals (XY, Z, parametric drive) generated at room temperature 
        are progressively attenuated and filtered at different temperature stages of the dilution refrigerator 
        before reaching the quantum processor at $\sim$10\,mK. Readout signals from the device are amplified by a 
        near-quantum-limited Josephson Parametric Amplifier (JPA) at the base temperature, followed by a HEMT 
        amplifier at 4\,K and further amplification at room temperature before demodulation and acquisition.}
        \label{fig:setup}
    \end{figure}

    Our measurement circuit is depicted in \autoref{fig:setup}. The quantum processor is mounted in an aluminium 
    sample holder at a base temperature of 10mK in a dilution refrigerator, protected with a magnetic 
    shielding and an infrared shielding. Full qubit control is implemented using four synchronized 
    Arbitrary Waveform Generators (AWGs, Tek5014C), which supply six pairs of sideband-modulated 
    signals for XY and readout control. These XY and readout control signals are generated from a 
    single microwave source, modulated at different sideband frequencies. One analog channel of an 
    AWG is dedicated to generating the parametric modulation pulse and Z control for the common qubit. 
    This method of control guarantees stable phase differences during the quantum tomography experiments. 
    To minimize phase errors and leakage to higher transmon levels, we employ Derivative Removal Adiabatic 
    Gate (DRAG) pulses for pulse correction and single-qubit rotations \cite{motzoi2009simple}. 

    For readout, six cavities are coupled to two separate transmission lines, each connected to a 
    Josephson Parametric Amplifier (JPA) \cite{kamal2009signal,murch2013observing}. 
    The JPA is pumped and biased by a signal generator 
    and a DC voltage source (YOKOGAWA GS210), providing approximately 20\,dB of gain with a bandwidth 
    of about 300\,MHz at 10\,mK. It serves as the first amplification stage, followed by a high-electron 
    mobility transistor amplifier at 4\,K and additional amplifiers at room temperature. This configuration 
    enables high-fidelity, simultaneous single-shot readout of all qubits. The JPA circuit design employs 
    standard 50\,$\Omega$ impedance matching without further impedance engineering. The detailed measurement and 
    readout techniques can be found in Ref.\cite{han2020error,cai2021impact,wang2022information}

\section{Device parameters}
    Our quantum processor comprises six transmon qubits ($Q_0$-$Q_5$) coupled to a shared bus 
    resonator, as shown in \autoref{fig:Fig1}(d) in the main text. We use four qubits ($Q_0$-$Q_3$) to implement the global 
    entangling gate. $Q_1$ and $Q_2$ are fixed frequency qubits, while $Q_0$ and $Q_3$ have their individual flux line 
    for frequency tuning. Each qubit is coupled to an independent Purcell-filtered cavity for individual 
    readout. The readout signals for qubits $Q_0$-$Q_2$ are multiplexed onto one transmission line, while those 
    for qubits $Q_3$-$Q_5$ are multiplexed onto a second, separate line. 

    We measure each qubit's energy relaxation time ($T_1$), Ramsey dephasing time ($T_2$), and echo dephasing 
    time ($T_{2}^{Echo}$), as well as the inter-qubit couplings. The resulting device parameters are summarized in 
    \autoref{tab:device_parameters}. 

    The coupling between higher energy levels of qubits gives rise to a cross-Kerr term 
    $\xi_{ZZ}a_1^\dagger a_1a_2^\dagger a_2$, resulting in a static ZZ interaction 
    \cite{han2020error,mundada2019suppression,kounalakis2018tuneable}. 
    We define $\xi_{ZZ}=\hbar(\omega_{11}-\omega_{01}-\omega_{10})$ as the static ZZ coupling strength 
    with respect to performance of single-qubit 
    gates. We measure the static ZZ interaction by a Ramsey-type measurement, which involves probing the 
    frequency shift of $Q_1$ with initializing $Q_2$ in either its ground or excited state. 

    We measure the coupling strength between each qubit and the bus by probing the qubit frequency shift 
    induced by photon number splitting. A pump pulse is sent to the bus to displace the photon state 
    followed by a probe pulse on the qubit to detect its frequency. We obtain qubit spectrum with coherent
    bus drive at different average cavity occupations (n). \autoref{fig:number_splitting} shows a representative qubit spectrum for $Q_3$. To improve the spectral 
    resolution, a long probe pulse of 5000 ns is used. The qubit response is observed at frequencies 
    $\omega_q + 2n\chi$ ($n$ = 0,1,2...) corresponding to the bus being in the Fock state $\ket{n}$ 
    \cite{xiong2025scalable,schuster2007circuit}. The coupling
    strength $h_j$ is extracted by $\chi=-\frac{h_j^2\alpha}{\Delta_{j,b}(\Delta_{j,b}-\alpha)}$ \cite{schuster2007circuit}, where $\alpha_j$ is the anharmonicity of the qubit, and $\Delta_{j,b}=\omega_q-\omega_b$.

    \begin{figure}[!t]
        \centering
        \includegraphics[width=0.6\textwidth]{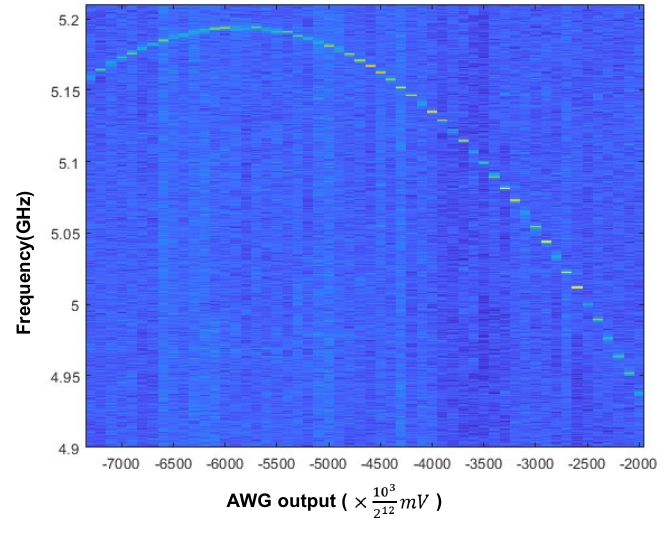}
        \caption{ \textbf{Flux map of the common qubit $Q_0$}. The transition frequency of $Q_0$ is plotted as a function of the 
        DC voltage bias applied to its dedicated flux line via an Arbitrary Waveform Generator (AWG). The characteristic 
        parabolic dependence is visible, with the maximum frequency corresponding to the flux-insensitive "sweet spot". 
        This map is used to precisely calibrate the DC bias and AC modulation amplitudes required for parametric gate operations.}
        \label{fig:spectrum}
    \end{figure}

    \begin{figure}[!ht]
        \centering
        \includegraphics[width=0.6\textwidth]{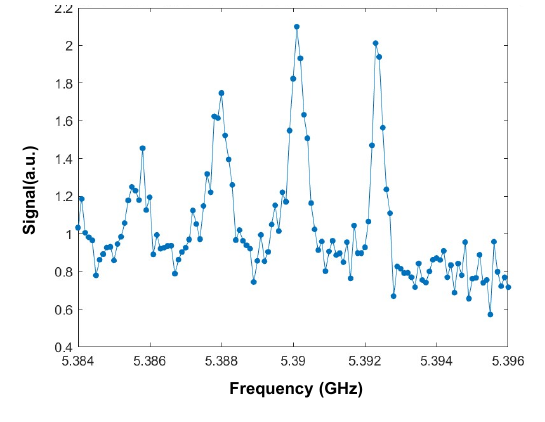}
        \caption{\textbf{Resonator-induced photon number splitting of $Q_0$}. The spectrum of qubit $Q_3$ is measured while the shared 
        bus resonator is driven with a coherent tone. The qubit's transition frequency splits into distinct peaks, with each 
        peak corresponding to a Fock state $\ket{n}$ of the resonator. This measurement confirms the dispersive 
        coupling between the qubit and resonator and allows for extraction of the dispersive shift $\chi$, which is then used
        to calculate the coupling strength $h_j$.}
        \label{fig:number_splitting}
    \end{figure}

\section{Theory}    
    In this section, we detail the theoretical derivation of the effective Hamiltonian that governs the parametric 
    interaction in our system. We begin by considering a system comprising $n$ fixed-frequency transmon qubits coupled 
    to a frequency-tunable transmon via a shared bus resonator. The full system Hamiltonian in the laboratory frame can be written as:

    \begin{equation}\label{eq:Ham_origin}
        \begin{aligned}
            H=& \sum_{j=1}^n (\omega_j a_j^\dagger a_j+\frac{\alpha_j}{2}a_j^\dagger a_j^\dagger a_j a_j)+
                (\omega_0 a_0^\dagger a_0+\frac{\alpha_0}{2}a_0^\dagger a_0^\dagger a_0 a_0)\\
                &+\sum_{j=1}^n g_j (a_0 a_j^\dagger + a_j a_0^\dagger)\\
            =& H_q+H_a+H_c
        \end{aligned}
    \end{equation}
  
    where $a_j^\dagger$ and $a_j$ are the creation and annihilation operators for the $j$-th qubit, $\omega_j$ is 
    its angular frequency, and $\alpha_j$ is its anharmonicity. The subscript $j=0$ denotes the frequency tunable qubit (common qubit). 
    The final term describes the exchange-type interaction between the tunable qubit and the other qubits. 
    This effective coupling strength, $g_j$, arises from a virtual photon exchange mediated by the bus resonator. 
    In the dispersive regime, where the qubit-resonator detunings are large compared to their coupling strengths, 
    a second-order Schrieffer-Wolff transformation yields 
    $g_j \approx \frac{h_0 h_j}{2}(\frac{1}{\omega_0 - \omega_b} + \frac{1}{\omega_j - \omega_b})$, where $h_j$ is the coupling strength 
    between the qubit $Q_j$ and the bus resonator, and $\omega_b$ is the frequency of the bus resonator. 
    
    To simplify the analysis by removing the fast-oscillating free-evolution terms, we transition to an interaction 
    picture defined by the free Hamiltonian of the qubits, $H_{\mathrm{free}} = \sum_{j=0}^n \omega_j a_j^\dagger a_j$. 
    The transformation is given by $U_1 = \exp(-it H_{\mathrm{free}})$, which yields the Hamiltonian:

    \begin{equation}\label{eq:Ham_rotating}
            H=\sum_{j=0}^n \frac{\alpha_j}{2}a_j^\dagger a_j^\dagger a_j a_j+
            \sum_{j=1}^n g_j (a_0 a_j^\dagger \mathrm{e}^{i\Delta_j t} + a_j a_0^\dagger \mathrm{e}^{-i\Delta_j t})
    \end{equation}

    where $\Delta_j=\omega_j-\omega_0$ is the detuning between the $j$-th qubit and the tunable qubit.
    Our parametric modulation is realized by applying an oscillating flux bias to the tunable qubit, which modulates 
    its transition frequency, $\omega_0(t)$. This introduces a time-dependent driving term to the Hamiltonian. 
    Because this term is diagonal in the qubit basis (proportional to $a_0^\dagger a_0$), it commutes with the 
    transformation $U_1$ and can be added directly to the Hamiltonian in Equation~\eqref{eq:Ham_rotating}. 
    Assuming a separate drive tone for each targeted interaction, the Hamiltonian becomes:

    \begin{equation}\label{eq:Ham_driving}
        \begin{aligned}
            H=&\sum_{j=0}^n \frac{\alpha_j}{2}a_j^\dagger a_j^\dagger a_j a_j+
            \sum_{j=1}^n g_j (a_0 a_j^\dagger \mathrm{e}^{i\Delta_j t} + a_j a_0^\dagger \mathrm{e}^{-i\Delta_j t})\\
            &+\sum_{j=1}^{n}\Omega_j \sin {(\nu_j t+\phi_j)}a_0^\dagger a_0
        \end{aligned}
    \end{equation}

    with $\Omega_j$ and $\nu_j$ being the amplitude and frequency of the flux modulation, respectively. To absorb the 
    driving term into the coupling part of the Hamiltonian, we move into a second interaction picture defined 
    by the transformation $U_2 = \exp\left(-i\int H_{\mathrm{drive}}(t') dt'\right)$:
    \begin{equation}
        \begin{aligned}
            U_2&=\mathrm{Exp}\left(-i\int \sum_{j=1}^{n}\Omega_j \sin {(\nu_j t+\phi_j)}a_0^\dagger a_0 dt\right)\\
                &=\mathrm{Exp}\left(i\sum_{j=1}^{n}\epsilon_j \cos{(\nu_j t+\phi_j)}a_0^\dagger a_0\right)
        \end{aligned}
    \end{equation}
    with $\epsilon_j = \Omega_j/\nu_j$ being the dimensionless modulation amplitude. After applying this transformation 
    and truncating the Hilbert space to the qubit manifold (thereby ignoring effects related to higher energy levels 
    and the anharmonicity terms $\alpha_j$), the Hamiltonian becomes:
    \begin{equation}\label{eq:Ham_interaction}
            H=\sum_{j=1}^n g_j (a_0 a_j^\dagger \mathrm{e}^{i\Delta_j t}\mathrm{e}^{i\xi}
            +a_j a_0^\dagger \mathrm{e}^{-i\Delta_j t}\mathrm{e}^{-i\xi})
    \end{equation}
    where for simplicity we define $\xi=\sum_{j=1}^{n}\epsilon_j \cos{(\nu_j t+\phi_j)}$.
    
    The time dependence in the exponents can be eliminated by choosing the modulation frequency 
    for each interaction to be resonant with the qubit detuning, i.e., $\nu_j = \Delta_j$. We 
    now apply the Jacobi-Anger expansion to the exponential terms: 
    $e^{i\epsilon \cos(\theta)} = \sum_{k=-\infty}^{\infty} i^k J_k(\epsilon) e^{ik\theta}$. 
    Substituting this into Equation~\eqref{eq:Ham_interaction} produces terms that oscillate at 
    various harmonics of the drive frequencies. By invoking the rotating-wave approximation (RWA), 
    we neglect all rapidly oscillating terms and retain only the time-independent (resonant) components. 
    This yields the final effective Hamiltonian:

    \begin{equation}\label{eq:Ham_effective}
            H=\sum_{j=1}^n g_j J_1(\epsilon_j)\Pi_{k\neq j}J_0(\epsilon_k) (i a_0 a_j^\dagger\mathrm{e}^{i\phi_j}
            +h.c.)
    \end{equation}
    This coupling Hamiltonian in \autoref{eq:Ham_effective} can realize an iSWAP-like entangling gate between 
    the tunable qubit and each of the other qubits when choosing proper modulation amplitude and time duration.

    This effective Hamiltonian describes a controllable exchange interaction with an effective coupling strength 
    $g_{j,\text{eff}} = g_j J_1(\epsilon_j) \prod_{k\neq j} J_0(\epsilon_k)$, where $J_n$ are Bessel functions of
     the first kind. The product term reflects the AC Stark shift induced by the other simultaneous drive tones. 
     By tuning the modulation amplitudes $\epsilon_j$ and the interaction time $t$, we can engineer a desired 
     unitary evolution. Specifically, for a two-qubit interaction, setting the initial phase of the drive to 
     $\phi_j = -\pi/2$ transforms the Hamiltonian into the form $H_{\text{eff}} = g_{j,\text{eff}} (a_0 a_j^\dagger + a_j a_0^\dagger)$. 
     An evolution under this Hamiltonian for a time $t = \pi/(4g_{j,\text{eff}})$ ideally 
     generates the $\sqrt{\mathrm{iSWAP}}$ gate, which is the subject of our experimental investigation.

    In practice, the phase of this gate requires careful tuning. The additional phase accumulated by the qubits 
    from their free evolution, which is tracked in the rotating frame defined by the transformation $U_1$, can be 
    compensated by applying single-qubit virtual $z$-rotations. However, applying such rotations is mathematically 
    equivalent to shifting the temporal origin of the entire gate sequence. Therefore, the essential tuning parameter 
    for the interaction basis itself is the initial phase, $\phi_j$, of the parametric pulse. We optimize this parameter 
    directly and, for the first gate in a sequence, set it to $\phi_j = -\pi/2$ to realize the desired $\sqrt{\mathrm{iSWAP}}$ 
    operation. For subsequent gates in a sequence, this phase must be advanced accordingly to account for the 
    evolution that occurred during the preceding gates.

\section{Calibration of parametric drive amplitude and frequency}
    In our experiment, we measure the simultaneous quantum-state population of all qubits as a function of 
    gate length and drive frequency by applying a parametric modulation pulse on the common qubit. This 
    parametric flux modulation induces a DC shift in the common qubit's frequency, thereby shifting the 
    resonant exchange frequency from its un-driven value, $\nu_{j,\Omega_j=0}$. Relative to the detuning between 
    each computational qubit and the common qubit, we observe drive frequency offsets of 8.3\,MHz, 7.6\,MHz, 
    and 10.8\,MHz for the two-, three-, and four-qubit cases, respectively. These offsets vary with the 
    amplitude of the parametric drive. At the offset driving frequency, coherent exchange between qubits 
    exhibits slower oscillation but higher inter-qubit exchange. Deviating from this frequency results in 
    faster oscillation and degraded quantum exchange, as shown in the Chevron pattern (\autoref{fig:Fig2}(e)) as an 
    example.
    
\begin{figure}[!ht]
        \centering
        \includegraphics[width=0.6\textwidth]{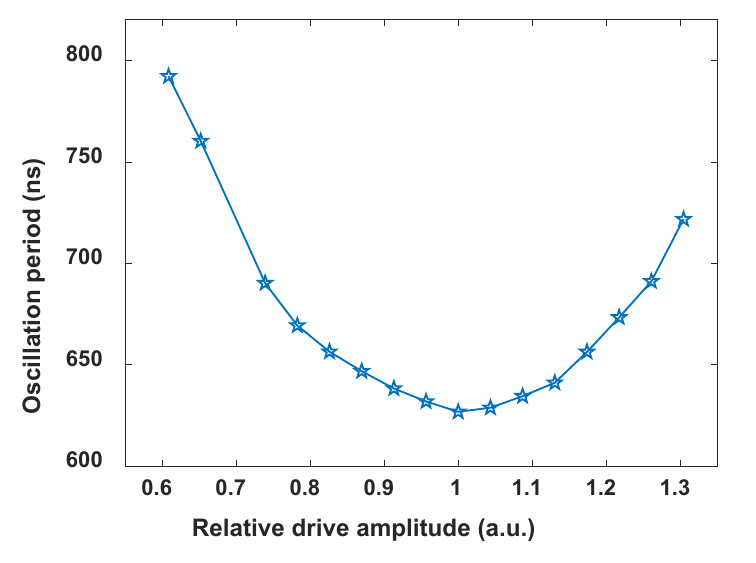}
        \caption{\textbf{Calibration of the parametric drive amplitude}. The oscillation period of a full population swap 
        between $Q_0$ and $Q_1$ is plotted as a function of the relative amplitude of the parametric drive. 
        The period is inversely proportional to the effective coupling strength $g_{eff}$. The optimal drive amplitude is 
        chosen at the minimum of the oscillation period, which corresponds to the maximum coupling strength and fastest gate speed.}
        \label{fig:time_vs_amp}
    \end{figure}

    We tune both the parametric drive amplitude and frequency to find a condition, which yields a maximum 
    inter-qubit exchange with the slowest exchange oscillation. Accordingly, we define the “oscillation 
    period” as the period of exchange oscillation (a full population swap) along the cut line corresponding to the offset drive 
    frequency. This period changes with drive amplitude. In practice, we select a drive amplitude with 
    corresponding frequency that give the shortest oscillation period as a reference driving power. \autoref{fig:time_vs_amp}
    shows the variation of oscillation period with driving power relative to the reference. The data 
    indicate that the oscillation period initially decreases with increasing drive power, reaches a minimum, 
    and then increases again.

\section{XEB}
    To evaluate the state preparation and measurement (SPAM) error-free fidelity of the $\sqrt{\mathrm{iSWAP}}$ 
    gate, which is not an element of the Clifford group, we perform cross-entropy benchmarking (XEB).
    The XEB protocol involves applying sequences of random single-qubit rotations chosen from a set of 64 distinct single-qubit gates. Each 
    	gate is a combination of a $\frac{\pi}{2}$ rotation around an axis in the xy-plane at $\frac{n\pi}{8}$ and a Z-rotation of $\frac{n\pi}{8}$, where $n=0, 1, \dots, 7$. 
    By measuring the output bitstrings $\{x_i\}$ for each gate sequence, we calculate the standard XEB fidelity, $F_{\mathrm{XEB}}$, as defined in Equation~\ref{eq:XEB_fidelity}, 
    where $p(x_i)$ is the probability of observing the bitstring $x_i$.

    \begin{subequations}\label{eq:XEB_fidelity}
        \begin{align}
            F_{\mathrm{XEB}}&=\frac{H(p_{uni},p_{th})-H(p_{exp},p_{th})}{H_(p_{uni},p_{th})-H_(p_{th},p_{th})} \\
            H(p,q)&=-\sum_{\{x_i\}} p(x_i)\log_2 q(x_i)
        \end{align}
    \end{subequations}

    During the XEB experiment, we observe that the gate fidelity is highly sensitive to phase 
    errors in the parametric pulse. This pulse is synthesized based on the pre-calibrated qubit 
    spectrum, an approach that implicitly assumes an ideal, instantaneous response from the flux 
    control line. However, non-negligible distortion and delay within the control system introduce 
    additional errors to the  $\sqrt{\mathrm{iSWAP}}$ gate. As these errors accumulate 
    throughout the duration of the two-qubit gate, they cannot be compensated by subsequent single-qubit operations.
    Therefore, a precise calibration of the qubit's $z$-control line response is crucial.

    To this end, we employ a Cryoscope protocol to characterize the actual flux, $\Phi_Q$, 
    at the chip level\cite{rol2020time}. The Cryoscope is a Ramsey-like experiment wherein square flux pulses 
    of varying lengths, but constant amplitude, are embedded between two $\frac{\pi}{2}$ pulses. 
    By measuring the accumulated qubit phase, $\phi$, induced by the off-idle pulse, we 
    can determine the on-chip frequency detuning via the relation $\frac{\Delta \phi}{\Delta t}$. 
    This frequency detuning is then converted to a flux value using the pre-calibrated 
    qubit spectrum, yielding response data with a temporal resolution set by the
    time domain resolution of the flux pulse.
    
   Informed by the frequency response data, we design a set of digital pre-distortion filters to 
   compensate for the phase errors. For long-timescale distortions (>10\,ns), 
   such as overshoot, undershoot, and slow drift of the flux pulse, we implement infinite impulse 
   response (IIR) filters. For short-timescale effects (<10\,ns), including fast oscillations 
   arising from the limited bandwidth of the control system and residual errors from the IIR filters, 
   we utilize finite impulse response (FIR) filters. The efficiency of these filters is verified by 
   routing the pre-distorted flux pulses back into the Cryoscope experiment. Both IIR and FIR 
   filters tend to temporally extend the pulse on the order of the overshoot/undershoot; 
   we mitigate this effect by introducing additional idle time before and after the flux pulse.

    Three degrees of freedom are available for phase tuning during the XEB experiment: virtual $z$-rotations
    applied to each qubit following the $\sqrt{\mathrm{iSWAP}}$ gate, and the initial phase of the parametric pulse. 
    The first two are straightforward, as our parametric pulse ideally generates a $\sqrt{\mathrm{iSWAP}}$ gate up 
    to single-qubit rotations. The third degree of freedom corresponds to the initial phase of the parametric drive in 
    the Hamiltonian. Several considerations are pertinent to this phase calibration. First, the phase of our 
    parametric pulse must compensate for the phase accumulated due to the detuning of the two 
    qubits from their physical interaction idle frequency point. The frequency of the applied pulse, however, 
    is slightly different from the qubit detuning. The dynamical part of the initial phase depends 
    solely on the qubit detuning, not the applied frequency. Second, the initial phase is related 
    to the pulse's temporal ordering, since virtual $z$-rotations after the gate are equivalent to 
    choosing a distinct temporal starting point. Consequently, we must adjust the initial phase of 
    each parametric pulse in addition to the dynamical one. Through the comprehensive calibration and optimization procedures described above, we achieve 
    an average XEB fidelity of $94.54(38)\%$ for the $\sqrt{\mathrm{iSWAP}}$ gate. We attribute the primary source 
    of infidelity to the decoherence error. 

\section{Cryoscope measurement}
    The fidelity of parametrically activated gates is critically dependent on the precise, time-domain control 
    of the qubit frequency. Non-ideal components in the control electronics, such as amplifiers and cables, 
    inevitably introduce distortions to the flux pulses, leading to coherent control errors. To mitigate these 
    effects, we first characterize the on-chip pulse shape using a Cryoscope protocol and then apply digital 
    pre-distortion filters to the input waveform \cite{rol2020time}.

    As described in the main text, the Cryoscope protocol is a Ramsey-like experiment designed to measure the time-resolved 
    frequency shift induced by a flux pulse. The sequence consists of two $\frac{\pi}{2}$ pulses separated by a 
    fixed time interval, with a square flux pulse of variable duration, $\tau$, embedded within this interval. 
    By sweeping the pulse duration $\tau$ and measuring the expectation values $\langle \sigma_x \rangle$ and 
    $\langle \sigma_y \rangle$ of the qubit after the final $\frac{\pi}{2}$ pulse, we reconstruct the total 
    accumulated phase, $\phi(\tau)$.

    The raw phase data contains both the desired signal from the pulse-induced detuning and high-frequency 
    oscillations. To isolate the former, we perform a series of post-processing steps. First, a Fourier 
    transform is applied to the phase data, $\phi(\tau)$, to identify the dominant oscillation frequency, 
    which corresponds to the static detuning of the qubit from its idle point. This frequency is used to demodulate 
    the signal, effectively removing the large-scale Ramsey fringes and highlighting the subtle deviations caused 
    by pulse distortions. Subsequently, a second-order Savitzky-Golay filter is applied to the demodulated phase 
    data. This smooths the signal while preserving its features and allows for a robust numerical calculation of 
    the derivative, $\frac{d\phi}{d\tau}$, which is directly proportional to the instantaneous frequency detuning. 
    Finally, using a pre-calibrated qubit flux-frequency map, we convert this time-resolved frequency detuning into 
    the effective on-chip flux pulse shape, $\Phi_Q(\tau)$.

    With the measured pulse response, we design digital filters to pre-distort the input waveform such that the 
    on-chip pulse becomes as close to the ideal square pulse as possible. We employ a combination of infinite 
    impulse response (IIR) and finite impulse response (FIR) filters, which can both be described by the general 
    linear difference equation:
    \begin{equation}
        a_0y[n] = \sum_{i=0}^{N} b_i x[n-i] - \sum_{j=1}^{M} a_j y[n-j]
    \end{equation}
    where $x[n]$ is the discrete input signal and $y[n]$ is the discrete output signal 
    (pre-distorted pulse). FIR filters are non-recursive ($a_j=0$ for $j\geq1$), whereas IIR filters are recursive.

    We use IIR filters to compensate for long-timescale distortions (>10\,ns), such as exponential undershoots, 
    which can be modeled as a single-pole decay. For a step function response $\eta(t)$ distorted by an exponential 
    undershoot of the form $s(t) = \eta(t)(1+A e^{-t/\tau_{\mathrm{IIR}}})$, the corresponding coefficients for a 
    discrete-time IIR filter that inverts this response can be analytically derived:
    \begin{subequations}
        \begin{align}
            a_0 &= 1 \\
            a_1 &= \alpha - 1 \\
            b_0 &= k\alpha - k + 1 \\
            b_1 &= -(\alpha - 1)(k - 1)
        \end{align}
    \end{subequations}
    where $\alpha = 1 - \exp(1/(f_s\tau_{\mathrm{IIR}}(A+1)))$, $k = -1/((A+1)(\alpha-1))$ for an undershoot ($A<0$), 
    and $f_s$ is the sampling rate of our arbitrary waveform generator.

    We use FIR filters to correct for short-time
    scale (<10\,ns) distortions, such as fast oscillations (ringing) from 
    limited bandwidth and any residual errors introduced by the IIR filter. Lacking a simple analytical model for these 
    distortions, we numerically optimize the FIR filter coefficients $\{b_i\}$. We employ the Covariance Matrix 
    Adaptation Evolution Strategy (CMA-ES) algorithm to find the 7 filter taps that minimize the mean squared 
    error between the ideal square pulse and the experimentally measured on-chip response after filtering.
 
\begin{figure}[!t]
        \centering
        \includegraphics[width=0.8\textwidth]{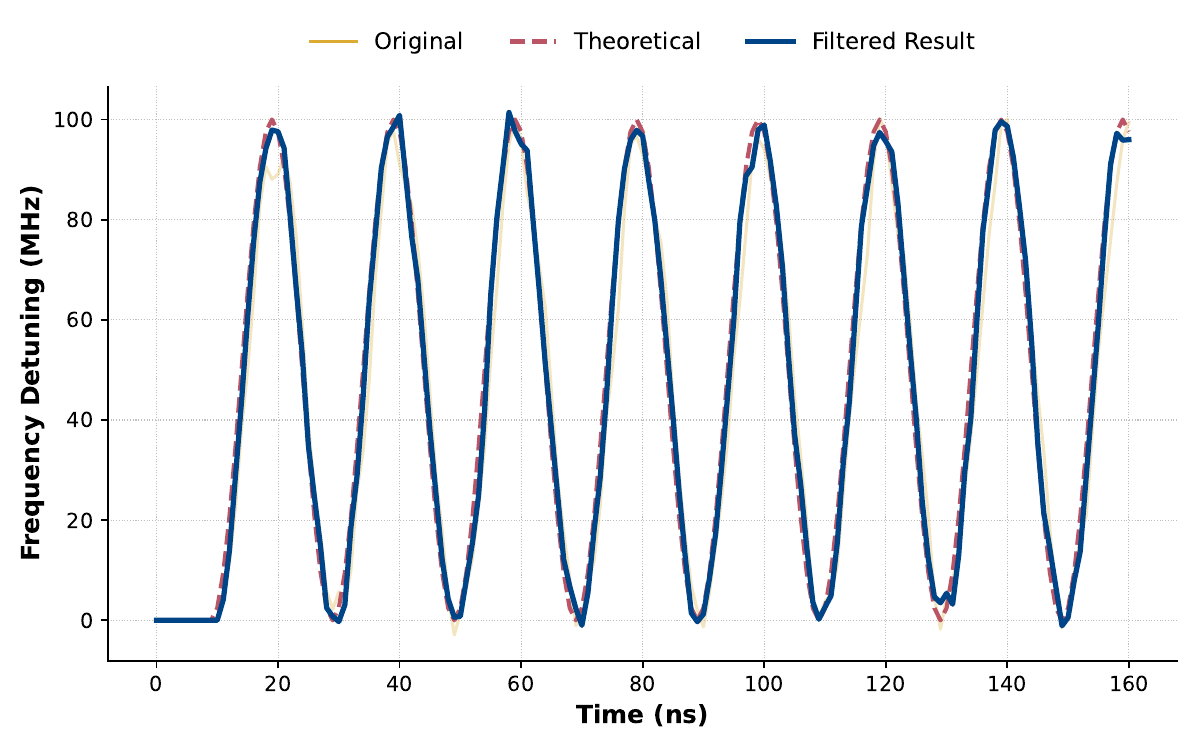}
        \caption{\textbf{Verification of pulse shaping for sinusoidal modulation}. Cryoscope measurement of the on-chip response to a 
        sinusoidal flux drive. The target waveform, a 50 MHz sinusoid, is shown as a dashed red line. The measured response 
        to this input pulse after correction by the IIR/FIR pre-distortion filters is shown in blue. The excellent agreement 
        demonstrates the filter's efficacy in producing high-fidelity, time-varying frequency modulation at the chip level.}
        \label{fig:2_cryoscope}
    \end{figure}

    The efficiency of these filters is verified by routing the pre-distorted flux pulses back 
    into the Cryoscope experiment. We choose the same DC flux bias as in the two-qubit gate experiment, applying parametric
    pulses with an amplitude corresponding to a 50\,MHz frequency shift($f_{pp}=$100\,MHz) and 50\,MHz in time domain.
    The Cryoscope results before and after the pre-distortion are shown in \autoref{fig:2_cryoscope}.

\section{QST data for $Q_0$-$Q_2$}

\begin{figure}[!ht]
	\centering
	\includegraphics[width=0.6\textwidth]{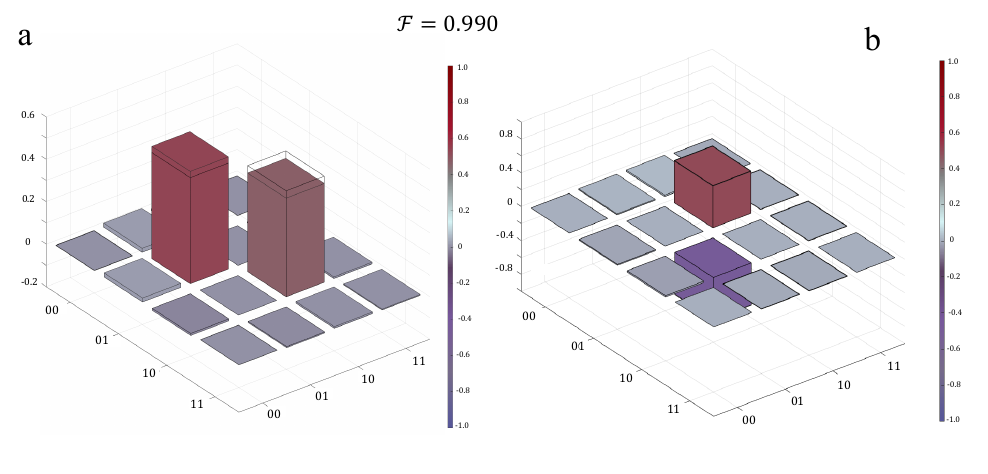}
	\caption{\textbf{QST for $Q_0$-$Q_2$}. Real part (a) and imaginary part (b) of a reconstructed 2-qubit density matrix for $Q_0$-$Q_2$ 
		(F = 0.990(2), $t_{gate}=365\,\mathrm{ns}$). Experimental results (colored bars) are compared to the ideal 
		state (transparent bars).}
	\label{fig:02ST}
\end{figure}

\autoref{fig:02ST} (a) and (b) show a representative two-qubit state density matrix for $Q_0$-$Q_2$.

\section{Decoherence error}
    Given the gate duration of 320-720\,ns, decoherence represents a primary source of infidelity for 
    the $\sqrt{\mathrm{iSWAP}}$ operation. In this section, we develop a theoretical model to quantify the impact 
    of environmental noise on the gate fidelity. We model this decoherence by considering two independent noise 
    channels acting on the qubits: energy relaxation (amplitude damping) and pure dephasing (phase damping).

    The effect of these noise processes is quantified using the Kraus operator formalism. 
    The average fidelity of a noise channel, which describes how well the channel preserves an arbitrary 
    input state on average, is given by\cite{li2024realization}:
    \begin{equation}\label{eq:fidelity_decoherence}
        F = \frac{d + \sum_k |\mathrm{Tr}(E_k)|^2}{d(d+1)}
    \end{equation}
    where $d$ is the dimension of the Hilbert space and $\{E_k\}$ are the Kraus operators that define the noise channel, 
    satisfying the completeness relation $\sum_k E_k^\dagger E_k = I$. Note that we have simplified the trace 
    term using this relation.

    For a single qubit, we define the probabilities of a relaxation or dephasing event occurring during the gate time $t$ as:
    \begin{align*}
        \text{Amplitude Damping Probability:} \quad & p_1 = 1 - e^{-t/T_1} \\
        \text{Phase Damping Probability:} \quad & p_\phi = \frac{1}{2}(1 - e^{-t/T_\phi})
    \end{align*}
    where $T_1$ is the energy relaxation time and $T_\phi$ is the pure dephasing time. For gate times $t \ll T_1, T_\phi$, 
    we can use the first-order approximations $p_1 \approx t/T_1$ and $p_\phi \approx t/(2T_\phi)$. The Kraus operators for 
    these single-qubit channels are:
    \begin{subequations}\label{eq:Kraus_decoherence}
        \begin{align}
            \text{Amplitude Damping:}\quad & E_0=\begin{pmatrix} 1 & 0 \\ 0 & \sqrt{1-p_1} \end{pmatrix},\\ &E_1=\begin{pmatrix} 0 & \sqrt{p_1} \\ 0 & 0 \end{pmatrix} \\
            \text{Phase Damping:}\quad & E_2=\begin{pmatrix} \sqrt{1-p_\phi} & 0 \\ 0 & \sqrt{1-p_\phi} \end{pmatrix},\\ &E_3=\begin{pmatrix} \sqrt{p_\phi} & 0 \\ 0 & -\sqrt{p_\phi} \end{pmatrix}
        \end{align}
    \end{subequations}

    To model the decoherence on an $N$-qubit system, we assume that the noise acts independently on each qubit. The total 
    noise channel is thus a tensor product of single-qubit channels, where the non-trivial Kraus operators act on one qubit while 
    the identity operator acts on the others. By constructing the multi-qubit Kraus operators for each channel and substituting them into 
    \autoref{eq:fidelity_decoherence}, we derive the average fidelity for systems of 2, 3, and 4 qubits. The resulting fidelities, keeping only 
    the first-order terms in $t/T_1$ and $t/T_\phi$, are summarized in \autoref{tab:decoherence_error}.

    \begin{table}[!ht]%The best place to locate the table environment is directly after its first reference in text
        \caption{\label{tab:decoherence_error}%
           Average state fidelity under amplitude damping 
           and phase damping for different qubit numbers.
            }
        \begin{ruledtabular}
        \begin{tabular}{cccc}
        \textrm{Qubit number}&
        \textrm{2}&
        \textrm{3}&
        \textrm{4}\\
        \colrule
        Amplitude damping & $1-\frac{2t}{5T_1}$ & $1-\frac{4t}{9T_1}$ & $1-\frac{8t}{17T_1}$\\
        Phase damping & $1-\frac{2t}{5T_\phi}$ & $1-\frac{4t}{9T_\phi}$ & $1-\frac{8t}{17T_\phi}$\\
        \end{tabular}
        \end{ruledtabular}
    \end{table}

\section{Numerical simulations}
    Using Qutip in PYTHON \cite{johansson2012qutip}, we perform numerical simulations with the Hamiltonian in 
    \autoref{eq:Ham_simu}, which 
    accounts for higher-level states by truncating each qubit's energy spectrum to three levels. Here, 
    $\omega_j$ denotes the frequency of $Q_j$, $g_{0,j}$ is the coupling strength between $Q_j$ and $Q_0$, and 
    $g_{0,j}=\frac{h_0 h_j}{2}(\frac{1}{\Delta_{0,b}} +\frac{1}{\Delta_{j,b}})$, where $\Delta_{j,b}=\omega_j-\omega_b$. 
    The parameters $\omega_j$  and $\nu_j$  represent the 
    amplitude and frequency of the parametric drive applied to the qubit pair $Q_0$-$Q_j$, and $\alpha_j$ is the 
    anharmonicity of $Q_j$. The fifth and sixth terms correspond, respectively, to random flux noise, 
    denoted $S_\phi^{(random)}$, and the static ZZ interaction $\xi_{ZZ}^j$ between $Q_j$ and $Q_0$. 
    
    We model the environmental flux noise as a characteristic $1/f$ power spectral density (PSD), 
    given by $S_{\phi}(f) = A_{\Phi}/f$\cite{sung2021realization}. The noise amplitude, $A_{\Phi}$, is an experimentally determined parameter 
    extracted from independent Ramsey and echo T2 measurements on the common qubit. From this PSD, we can generate 
    representative time-domain noise traces.

    The impact of $1/f$ noise is dominated by its low-frequency components. For a gate of duration $t_{\text{gate}}$, 
    noise components with frequencies $f \ll 1/t_{\text{gate}}$ do not average out and act as a quasi-static
    frequency offset on the qubit. This offset is effectively constant during a single experimental shot but 
    varies from shot to shot. We therefore model this dominant noise effect by sampling a random frequency offset, 
    $\delta\omega$, from a Gaussian distribution with zero mean and a variance, $\sigma^2$, given by the integrated 
    noise power:
    \begin{equation}
        \sigma^2 = \int_{f_{\text{low}}}^{1/t_{\text{gate}}}  S_{\phi}(f) df
    \end{equation}
    The integration runs from a low-frequency cutoff of $f_{\text{low}} = 10^{-4}$\,Hz up to the frequency 
    corresponding to the gate's timescale (The high cutoff frequency is set to be about 3\,GHz. But for $1/f$ noise, the power at high frequencies 
    drops off so quickly that the contribution above 1-2\,GHz is negligible).

    The random flux 
    noise, $S_\phi^{(random)}$, is generated as a random waveform based on the experimentally extracted 
    flux-noise spectral density of $S_\phi=(2.45 \mu\phi_0)^2/\mathrm{Hz}$. Our simulations account for qubit decoherence, 
    dephasing, flux noise, and static ZZ interactions, with all simulation parameters set to their 
    experimentally measured values.    
    
    \begin{equation}\label{eq:Ham_simu}
        \begin{aligned}
        H=&\sum_{j=0}^{n}\omega_j a_j^\dagger a_j + 
            \sum_{j=0}^{n} \frac{\alpha_j}{2} a_j^\dagger a_j^\dagger a_j a_j + 
            \sum_{j=1}^{n} g_j(a_0a_j^\dagger \mathrm{e}^{i\Delta_jt+h.c.}) \\ &+
            \sum_{j=1}^{n} \Omega_j \sin(\nu_j t + \phi_j)a_0^\dagger a_0 +
            S_\phi^{(random)}(t)  a_0^\dagger a_0 +
            \sum_{j=1}^{n} \xi_{ZZ}^j a_0^\dagger a_0^\dagger a_0a_0 
        \end{aligned}
    \end{equation}

    The error infidelities arising from flux noise and ZZ interactions are extracted by comparing the 
    fidelity with and without the corresponding noise channel. We first simulate the time evolution of 
    the Hamiltonian under experimental conditions for systems of two, three, and four qubits. The 
    simulated qubit occupation probabilities during multi-qubit exchange oscillations are shown in 
    \autoref{fig:Fig3}(a). These simulation results are consistent with the experimental data.
    
    We next perform quantum state tomography (QST) and gate simulations to extract the error infidelity 
    contributed by each individual noise channel, summarizing the results in Table I in the main text. The state fidelity 
    is calculated with $F(\rho,\sigma)=\sqrt{\sqrt{\rho}\sigma\sqrt{\rho}}$  , where $\sigma$ is the density matrix 
    of the ideal final state and $\rho$ 
    is the simulated density matrix. The gate fidelity is then estimated by averaging the state fidelities 
    over the 36 two qubit initial states that are formed from the tensor product of the six standard single 
    qubit states, $\{\ket{0}, \ket{1}, \frac{1}{\sqrt{2}}(\ket{0} \pm \ket{1}), \frac{1}{\sqrt{2}}(\ket{0} \pm i\ket{1})\}^{\otimes 2}$. 
    We also simulate the evolution 
    of the gate unitary operator and extract the gate fidelity at the gate time when the entanglement 
    reaches its maximum. The simulation is performed by modeling each transmon as a three-level system (qutrit) to account for potential 
    leakage to the non-computational $\ket{2}$ state. We then perform a Monte Carlo estimation of the noise-induced 
    infidelity. For each run, a static frequency offset, sampled from the Gaussian distribution described above, 
    is added to the common qubit's Hamiltonian, and the full unitary evolution is simulated. By averaging the 
    process fidelity over 1000 such noise realizations, we determine the average infidelity contribution from 
    low-frequency flux noise.

    The results of this analysis, combined with the other simulated error channels, are summarized in 
    Table I in the main text. This comprehensive breakdown shows that the infidelity from flux noise, 
    $\epsilon_{\text{flux}}$, is on the order of $10^{-3}$ to $10^{-4}$, confirming that it is a sub-dominant 
    error source compared to qubit decoherence and coherent control errors for our current device.
    
    To verify the potential of our parametric global gate scheme, we simulate the generation of entangled 
    states in systems of two to six qubits using state of the art superconducting qubit parameters. The six 
    qubits $Q_0$-$Q_5$ have frequencies $\omega_j/2\pi = 5.0, 5.25, 5.18, 5.10, 5.06$, and 5.03\,GHz for $j=0\sim5$, 
    respectively, and the resonator frequency is $\omega_b/2\pi$=5.40\,GHz, with $Q_0$ serving as the common qubit. 
    We take into account qubit decoherence, dephasing, flux noise, and static ZZ interactions for the 
    simulation. We assign a relaxation time and a dephasing time of 60 $\mathrm{\mu s}$ to all qubits, representing 
    an achievable decoherence benchmark for current processors, and apply a static ZZ coupling 
    $\frac{\xi_{ZZ}^j}{2\pi}=2.5\times 10^4 \, \mathrm{Hz}$ between the common qubit and each computational qubit ($j=1\sim5$). The flux 
    noise spectral density is set to $S_\phi=(2.45 \mu\phi_0)^2/\mathrm{Hz}$, consistent with the values used in real device 
    simulations.
    
\begin{figure}[!ht]
        \centering
        \includegraphics[width=0.8\textwidth]{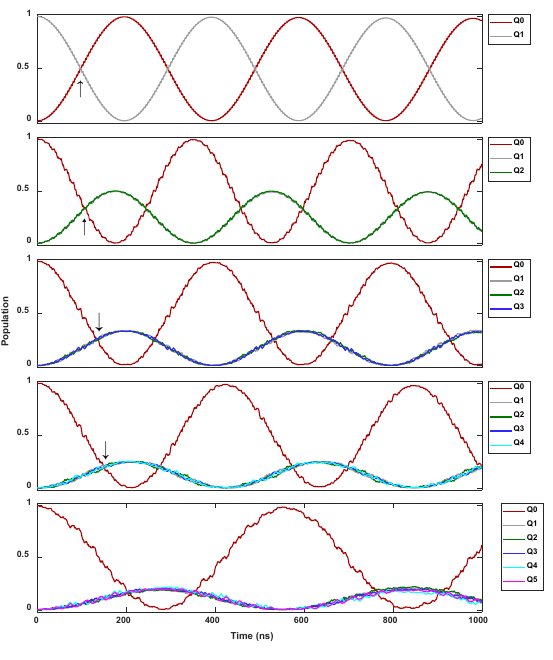}
        \caption{ \textbf{Simulated generation of multi-qubit W-states}. The population dynamics of systems from two to six qubits are 
        shown under a global parametric drive. The system is initialized in the state $\ket{10\cdots0}$. The drive induces coherent oscillations 
        that distribute the single excitation among all participating qubits. The black arrows indicate the time when the 
        system first reaches a state of maximum entanglement (an equal-superposition W-state), at which the state 
        fidelities are extracted. Simulation parameters: Bus resonator frequency: $\omega_b/2\pi$=5.40\,GHz; Qubit frequency: $\omega_j/2\pi = 5.0, 5.25, 5.18, 5.10, 5.06$, 5.03\,GHz,  
        $h_j/2\pi$=28\,MHz, $\alpha_j=-200 \,\mathrm{MHz}$, $\frac{\xi_{ZZ}^j}{2\pi}=2.5\times 10^4 \, \mathrm{Hz}$ for $j=0\sim5$; $T_1=60\, \mathrm{\mu s}$ and $T_2 Echo=60 \,\mathrm{\mu s}$; $S_\phi=2.45 \mu\phi_0/\sqrt{\mathrm{Hz}}$
        Two-qubit case: $\Omega_1/2\pi$=1200\,MHz; Three -qubit case: $\Omega_1/2\pi$=1200\,MHz, $\Omega_1$:$\Omega_2$=1:0.9; Four-qubit case: $\Omega_1/2\pi$=980\,MHz, $\Omega_1$:$\Omega_2$:$\Omega_3$=1:0.91:0.59;
        Five-qubit case: $\Omega_1/2\pi$=820\,MHz, $\Omega_1$:$\Omega_2$:$\Omega_3$:$\Omega_4$=1:0.89:0.6:0.36; Six-qubit case: $\Omega_1/2\pi$=1000\,MHz, $\Omega_1$:$\Omega_2$:$\Omega_3$:$\Omega_4$:$\Omega_5$=1:0.76:0.49:0.19:0.2.}  
        \label{fig:simulation}
    \end{figure}

    The resulting multi qubit exchange oscillations for systems of two to six qubits are shown in \autoref{fig:simulation}. 
    At the points of maximum entanglement (indicated by black arrows), we obtain high state fidelities of 
    99.91\%, 99.88\%, 99.85\%, 99.80\%, and 99.70\%, respectively.

\begin{table*}[!ht]
    \caption{\label{tab:device_parameters}%
        Device parameters
    }
    \begin{ruledtabular}
        \begin{tabular}{l c c c c c} 
        % I changed 'c' to 'l' (left align) for the first column, usually looks better
            \textrm{} & 
            $Q_0$ & 
            $Q_1$ & 
            $Q_2$ & 
            $Q_3$ & 
            Bus resonator \\
            \colrule
            $\omega_r/2\pi$(GHz) & 7.00075 & 7.0229 & 7.03875 & 7.0777 & 5.5208\\
            $\omega_j/2\pi$(GHz) & 5.0408\footnotemark[1] 5.1941\footnotemark[2] & 5.0992 & 5.2056 & 5.2347\footnotemark[1] 5.3928\footnotemark[2] & --\\
            $g_{r,j}/2\pi$(MHz) & 83.9 & 86.2 & 90.0 & 54.4 & --\\
            $T_1(\mathrm{\mu s})$ & 30.4\footnotemark[1] & 31.7 & 38.5 & 33.2\footnotemark[1] & --\\
            $T_2^{Ramsey}(\mathrm{\mu s})$   & 5.1\footnotemark[1] & 25.3 & 31.0 & 2.6\footnotemark[1] & --\\
            $T_2^{Echo}(\mathrm{\mu s})$    & 17.3\footnotemark[1] & 44.1 & 40.3 & 16.6\footnotemark[1] & --\\
            $g_j/2\pi$(MHz)   & 17.3 & 22.8 & 18.2 & 15.2 & -- \\
            $\xi_{ZZ}/2\pi$(kHz)   & -- & 26.2 & 30.1 & 28.0 & -- \\
            $\alpha_j$(MHz)   & -209.8 & -209.3 & -210.2 & -204.3 & -- \\
        \end{tabular}
    \end{ruledtabular}
    \footnotetext[1]{off sweetspot}
    \footnotetext[2]{on sweetspot}
\end{table*}

% \clearpage
% \bibliographystyle{plain}
% \bibliographystyle{unsrt}
% \bibliography{sup_ref}

\end{document}